\documentclass[amsmath,amssymb,twocolumn,amsfonts,aps,prb,10pt]{revtex4-1}
\usepackage{graphicx}
\usepackage[version=3]{mhchem}
\usepackage{tabularx}
\usepackage{datetime}
\usdate

\newcommand{\leftsub}[2]{{\vphantom{#2}}_{#1}{#2}}
\newcommand{\sfrac}[2]{\left.#1\middle/#2\right.}
\newcommand{\br}[1]{\left(#1\right)}
\newcommand{\fbr}[1]{\!\br{#1}}
\newcommand{\rconf}{\mathfrak{R}}

\newcommand{\overunder}[2]{\genfrac{}{}{0pt}{}{#1}{#2}}
\newcommand{\abs}[1]{\left|#1\right|}

\let\oldsqrt\sqrt
\def\sqrt{\mathpalette\DHLhksqrt}
\def\DHLhksqrt#1#2{%
\setbox0=\hbox{$#1\oldsqrt{#2}$}\dimen0=\ht0
\advance\dimen0-0.2\ht0
\setbox2=\hbox{\vrule height\ht0 depth -\dimen0}%
{\box0\lower0.4pt\box2}}

\begin{document}
\title{Elucidating Fermi's Golden Rule via bound-to-bound transitions in a confined hydrogen atom}
\author{L. M. Ugray}
\author{R. C. Shiell}
\email{ralphshiell@trentu.ca}
\affiliation{Trent University, Department of Physics and Astronomy, Peterborough, ON K9J 7B8, Canada}
\date{\today{}}

\begin{abstract}
We demonstrate an effective method for calculating bound-to-continuum cross-sections by examining transitions to bound states above the ionization energy that result from placing the system of interest within an infinite spherical well.  Using photoionization of the hydrogen atom as an example, we demonstrate convergence between this approach for a large volume of confinement and an exact analytical alternate approach that uses energy-normalized continuum wavefunctions, which helps to elucidate the implementation of Fermi's Golden Rule.  As the radius of confinement varies, the resulting changes in physical behavior of the system are presented and discussed.  The photoionization cross-sections from a variety of atomic states with principal quantum number $n$ are seen to obey particular scaling laws.
\end{abstract}

\maketitle
\section{Introduction}\label{sec:intro}

While undergraduate students usually encounter \emph{bound-to-bound} transitions,\cite{griffiths} they typically have little experience with understanding and performing calculations on transitions of a \emph{bound-to-continuum} nature, as exemplified by ionization and molecular dissociation.  Fermi's Golden Rule,\cite{sakurai}
\begin{equation}\label{eq:fermirule}
 \Gamma_{i\rightarrow \{j\}}=\frac{2\pi}{\hbar}\overline{\left|\left<j\right|\hat{H}'\left|i\right>\right|^2}\rho\fbr{W},
\end{equation}
provides an expression for the transition rate from an initial bound state $i$ to the infinitude of continuum states $\{j\}$ in a small energy range centered about $W$ with $\rho\fbr{W}$ being the density of states with the same symmetry as $j$.  While this is often presented\cite{fermi} for a time-independent perturbation, $\hat{H}'$, this rule can also apply for an sinusoidal perturbation, $\mathcal{\hat{H}}'\fbr{t}$, with angular frequency $\omega$, by setting $\hat{H}'=\mathcal{\hat{H}}'\fbr{t}\!/\!\br{e^{i\omega t}+e^{-i\omega t}}$, rendering $\hat{H}'$ again time-independent.\cite{atkins,starace,wavepacket} However, when applying Fermi's Golden Rule it is not immediately clear how to interpret the product of $\rho\fbr{W}$, which is infinite for states in the continuum where all energies are allowed, and the mean of the square of the matrix element which incorporates the space-normalized wavefunctions $\{\left<j\right|\}$, which have zero amplitude as they extend over all space.

In this paper we use two methods to resolve this mathematical dilemma, and in doing so provide a concrete demonstration of the consistency between them.  We thereby elucidate some of the subtleties of Fermi's Golden Rule which are often omitted in texts of introductory quantum mechanics,\cite{griffiths,liboff,bransden} but are important to understand for its applicability.  The first method uses a particular form of normalization for the continuum states called \emph{energy-normalization},\cite{cowan} which we introduce below.  The second method involves placing the system in an infinite well (\emph{box normalization})\cite{tellinghuisen} which causes all states to be bound so that states in the previously continuous regime become discretized, allowing the familiar techniques from a bound-to-bound approach to be used to perform bound-to-continuum calculations.

We use as an example photoionization of hydrogen, the process whereby the atom absorbs incident light such that the electron is no longer bound to the system.  Performing calculations related to a confined hydrogen atom is not a new concept,\cite{maize,wilcox,varandas,laughlin,capitelli,stevanovic,aquino} but past calculations of this kind have been done primarily to determine the properties of the atom inside a confining structure, and here we use confinement as a tool to determine the properties of the unconfined system.

This paper is organized as follows:  in Section \ref{sec:btb} we introduce the concept of a cross-section, apply it to more familiar bound-to-bound photoexcitations, and provide results specific to the hydrogen atom.  In Section \ref{sec:btc} we examine bound-to-continuum transitions, using an analytical approach for the unconfined atom, introduce energy normalization, and reproduce, for later comparison, exact calculations of photoionization cross-sections for hydrogen, the archetypal charged binary system.  In Section \ref{sec:wellH} we implement box-normalization, demonstrate convergence for increasing radius of confinement, and show that calculations using this method agree with the exact calculations from the previous section.

Two commonly encountered constants used throughout this paper are $\mathrm{a}_0$, the Bohr radius ($=\sfrac{4\pi\epsilon_0\hbar^2}{\br{m_ee^2}}\allowbreak\approx5.29{\times}10^{-11}\,\mathrm{m}$), and $\mathrm{Ry}$, the Rydberg unit of energy ($=m_ee^4/[\br{4\pi\epsilon_0}^22\hbar^2]\allowbreak\approx2.18{\times}10^{-18}\,\mathrm{J}$).

\section{Cross-sections for bound-to-bound transitions}\label{sec:btb}

A physical parameter that quantifies the absorption of light by a sample of atoms with some distribution of initial states is the \emph{cross-section}, $\sigma$.  The average photon absorption rate per atom due to an incident narrowband photon flux, $F$ (photons/unit time/unit area), is given\cite{bethe} by $\Gamma=\sigma\fbr{\omega}F$, and thus the cross-section indicates the average effective area of an atom presented to the light at angular frequency $\omega$. By considering the photon absorption rate of a single weak incident beam (i.e. assuming negligible population inversion) traveling in the positive $x$-direction within a thickness $dx$ of a sample, it can be shown that the intensity of the light, $I\fbr{x}$, obeys
\begin{equation}\label{eq:sigmadef}
 \frac{dI}{dx}=-\sigma\fbr{\omega} NI,
\end{equation}
where $N$ is the number density of all particles.  Eq.~\eqref{eq:sigmadef} is often given in textbooks as the definition for $\sigma$.

For bound-to-bound transitions between two particular quantum states, a standard semiclassical treatment using the time-dependent Schr\"odinger equation for a narrowband, weak incident light beam gives\citep{bernath}
\begin{equation}
 \Gamma_{i\rightarrow j}=\frac{\pi\omega}{c\epsilon_0 \hbar}\left|\left<j\right|\boldsymbol{\hat{d}}\left|i\right>\right|^2g\fbr{\omega-\omega_0}F_\epsilon,
\end{equation}
where $\boldsymbol{\hat{d}}$ is the electric dipole moment operator due to all charges in the atom, $F_\epsilon$ is the incident photon flux with polarization $\epsilon$ which couples the transition $i{\rightarrow}j$, and $g\fbr{\omega-\omega_0}$ is the lineshape of the transition with $\int\!g\,d\omega=1$.  In general, however, many atomic levels are degenerate and light can couple many of these states, so the average absorption rate per atom for an ensemble of atoms is given by
\begin{equation}
\Gamma=\sum\limits_{\{i,\,j\}}\Gamma_{i\rightarrow j}\sfrac{N_i}{N},
\end{equation}
where the summation runs over all pairs of states coupled by the incident light, and $\sfrac{N_i}{N}$ is the fraction of atoms in each initial state $\left|i\right>$.

It is usual to report absorption cross-sections that assume the atoms are evenly distributed among all $2J+1$ possible $M$ initial states.  This distribution is frequently encountered, and $\Gamma$ is then by symmetry independent of the polarization characteristics of the incident light.  For ease in deriving this cross-section, we will arbitrarily consider incident light which is linearly polarized along the $z$-axis ($\pi$-polarized, so $\Delta M{=}0$).  For a transition $\gamma J{\rightarrow}\gamma'J'$ this gives
\begin{equation}
\sigma=\frac{\pi\omega}{c\epsilon_0 \hbar}\frac{g\fbr{\omega-\omega_0}}{2J+1}\sum\limits_{M}\left|\left<\gamma'J'M\right|\hat{d}_z\left|\gamma JM\right>\right|^2,
\end{equation}
where $\hat{d}_z$ is the $z$-component of the electric dipole moment operator.  For the particular case of $n\ell{\rightarrow}n'\ell'$ transitions within the hydrogen atom, neglecting electron and nuclear spin, this gives
\begin{align}
 \begin{split}\label{eq:hydrogensigma}
  \sigma&=\frac{\pi\omega e^2}{c\epsilon_0 \hbar}\frac{g\fbr{\omega-\omega_0}}{2\ell+1}\left|D_{n\ell\rightarrow n'\ell'}\right|^2\\
  &\qquad\times\sum\limits_{m}\left|\int\limits_0^{2\pi}\!\int\limits_0^\pi {Y_{\ell'}^{m}}^* \br{\sqrt{\frac{4\pi}{3}}Y_1^0} Y_\ell^m\mathrm{sin}\theta\,d\theta\, d\phi\right|^2\\
  &=\frac{\pi\omega e^2}{c\epsilon_0 \hbar}\frac{g\fbr{\omega-\omega_0}}{2\ell+1}\left|D_{n\ell\rightarrow n'\ell'}\right|^2\sum\limits_{m}\frac{\ell_\mathrm{max}^2-m^2}{4\ell_\mathrm{max}^2-1}\\
  &=\frac{\pi\omega e^2}{3c\epsilon_0 \hbar}\frac{\ell_\mathrm{max}}{2\ell+1}\left|D_{n\ell\rightarrow n'\ell'}\right|^2g\fbr{\omega-\omega_0},
 \end{split}
\end{align}
where the angular integral was evaluated using Ref.~\citenum{zare}, $\ell_\mathrm{max}$ is the greater of $\ell$ and $\ell'$, $\abs{\ell-\ell'}=1$, and the radial matrix element is defined in terms of the well-known space-normalized radial wavefunctions, $R_{n\ell}$, by
\begin{equation}\label{eq:Dnlnl}
 D_{n\ell\rightarrow n'\ell'}=\int\limits_0^{\infty}R_{n'\ell'}^*r\,R_{n\ell}\,r^2\,dr.
\end{equation}

It follows that the \emph{integrated cross-section}, a commonly presented quantity that reflects the overall strength of a transition by integrating $\sigma$ over all incident frequencies, $\nu$, is  given by
\begin{equation}
 \int\!\!\sigma\,d\nu=\frac{\omega e^2}{6c\epsilon_0 \hbar}\frac{\ell_\mathrm{max}}{2\ell+1}\left|D_{n\ell\rightarrow n'\ell'}\right|^2.
\end{equation}

\section{Cross-sections for bound-to-continuum transitions: Energy-normalization}\label{sec:btc}

One approach to evaluating the problematic product described in Section~\ref{sec:intro} is to use energy-normalized wavefunctions for the continuum states, defined by:\citep{landau}
\begin{equation}\label{eq:energynormdef1}
 \int\limits_\mathrm{\overunder{all}{space}}\!\!\!\Psi_W^*\Psi_{W'}\, d\tau=\delta\fbr{W-W'}.
\end{equation}
Integrating over $W'$ on both sides for an arbitrarily small range centered about $W$ gives another representation of this equation:
\begin{equation}\label{eq:energynormdef}
 \int\limits_\mathrm{\overunder{all}{space}}\!\!\!\Psi_W^*\!\!\!\!\int\limits_{W-\Delta W}^{W+\Delta W}\!\!\!\!\Psi_{W'} dW' d\tau=1.
\end{equation}
We examine a hydrogen atom with space-normalized continuum wavefunctions $j=Y_{\ell}^{m}R_{W\ell}$ (i.e. a radial wavefunction with zero amplitude) and show that multiplication by $\sqrt{\rho\fbr{W}}$ (which is infinite) gives energy-normalized finite-amplitude wavefunctions satisfying  Eq.~\eqref{eq:energynormdef}.  The left hand side of this equation takes the form
\begin{align}
 \begin{split}
  &\int\!\!\!\!\!\!\int\limits_\mathrm{\overunder{all}{space}}\!\!\!\!\!\!\int\!\! \sqrt{\rho}\,{Y_{\ell}^{m}}^*R_{W\ell}^*\!\!\!\!\!\!\!\int\limits_{W-\Delta W}^{W+\Delta W}\!\!\!\!\!\!\!\! \sqrt{\rho}\,Y_{\ell}^{m}R_{W'\ell}\,dW'\,d\tau\\
  =&\int\limits_0^\infty r^2R_{W\ell}^*\!\!\!\int\limits_{W-\Delta W}^{W+\Delta W}\!\!\! R_{W'\ell}\frac{dn}{dW'}\,dW'\,dr\\
  =&\int\limits_0^\infty r^2R_{W\ell}^*\!\int\limits_{n^-}^{n^+}\! R_{W'\ell}\,dn\,dr=\int\limits_0^\infty r^2R_{W\ell}^*\!R_{W\ell}\,dr\equiv 1,
 \end{split}
\end{align}
where we have used in the first step that each $\rho$ is $\sfrac{dn}{dW}$ evaluated at $W$ and $W'$ respectively, with $W'$ bounded by $W\pm\Delta W$, and in the final step the integral over $n$ is unity since there is only one radial wavefunction $R_{W'\ell}$ which is not orthogonal to $R_{W\ell}$.

The photoionization cross-section for hydrogen can then be derived from Fermi's Golden Rule by including the density of states within the bra: $\left<\sqrt{\rho}\,j\right|$, and using for this energy-normalized continuum wavefunctions (energy \mbox{$W>0$}) with radial component:\cite{burgess}
\begin{align}\label{eq:normhydrogenabove}
 \begin{split}
  &R_{W\ell}^\mathrm{\,en}\fbr{r}=\sqrt{\frac{2\prod\limits_{s=0}^\ell \br{1+\frac{W}{\mathrm{Ry}}s^2}}{\br{1{-}e^{-\pi\sqrt{\frac{4\mathrm{Ry}}{W}}}}\mathrm{Ry}\,\mathrm{a}_0^3}}\,\frac{\br{\sfrac{2r}{\mathrm{a}_0}}^\ell}{\br{2\ell+1}!}\\
  &\quad\times e^{\frac{ir}{\mathrm{a}_0}\sqrt{\frac{W}{\mathrm{Ry}}}}\leftsub{1}{F}_1\fbr{\ell{+}1{-}i\sqrt{\frac{\mathrm{Ry}}{W}};\,2\ell{+}2;\,-2i\frac{r}{\mathrm{a}_0}\sqrt{\frac{W}{\mathrm{Ry}}}}.
 \end{split}
\end{align}
The normalization constants herein can be derived by applying Eq.~\eqref{eq:energynormdef} to the asymptotic form of the radial wavefunction, and these result in wavefunctions that when multiplied by $r$ tends to an oscillatory function with constant amplitude of $\br{\mathrm{a}_0^2\pi^2W\mathrm{Ry}}^{\sfrac{-1}{4}}$ for large $r$.\cite{tellinghuisen}  Note that for $W=-\sfrac{\mathrm{Ry}}{n^2}$ the functional form of Eq.~\eqref{eq:normhydrogenabove}, which contains \mbox{$\leftsub{1}{F}_1\br{a;\,b;\,z}$}, the confluent hypergeometric function,\cite{gradshteyn} reduces to that of the well-known generalized Laguerre polynomials for hydrogenic bound states.

Again assuming a uniform distribution of initial $m$ states and employing $\pi$-polarized light with electric field amplitude $E_0$, and identifying $\hat{H}'$ to be $\sfrac{d_z E_0}{2}$, Eq.~\eqref{eq:fermirule} becomes
\begin{align}
 \begin{split}
  &\Gamma_{n\ell\rightarrow W\ell}=\frac{1}{2\ell+1}\frac{2\pi}{\hbar}
%\\  &\quad\times  
  \sum\limits_m \left|\left<Y_{\ell'}^mR_{W\ell'}^\mathrm{\,en}\left|\frac{d_z E_0}{2}\right|Y_{\ell}^mR_{n\ell}\right>\right|^2.
 \end{split}
\end{align}  
By equating $E_0^2$ to $\sfrac{2\omega F_\pi\hbar}{c\epsilon_0}$, the photoionization cross-section can be seen to be:
\begin{align}
\begin{split}\label{eq:sigmacont}
  \sigma=&\frac{4\pi^2\omega \mathrm{a}_0}{c}\frac{2\mathrm{Ry}}{2\ell+1}\left|D_{n\ell\rightarrow W\ell'}\right|^2\times\\
  &\sum\limits_{m}\!\left|\int\limits_0^{2\pi}\!\int\limits_0^\pi\!{Y_{\ell'}^{m}}^*\!\br{\!\sqrt{\frac{4\pi}{3}}\!Y_1^0\!}\!Y_\ell^m\mathrm{sin}\theta\,d\theta\, d\phi\right|^2\\
  =&\frac{4\pi^2\omega \mathrm{a}_0 2\mathrm{Ry}}{3c}\frac{\ell_\mathrm{max}}{2\ell+1}\left|D_{n\ell\rightarrow W\ell'}\right|^2,
 \end{split}
\end{align}
where \mbox{$\hbar\omega=W+\sfrac{\mathrm{Ry}}{n^2}$.}  By analogy to Equation~\eqref{eq:Dnlnl}, we define
\begin{equation}\label{eq:DnlWl}
 D_{n\ell\rightarrow W\ell'}=\int\limits_0^{\infty}{R_{W\ell'}^\mathrm{\,en}}^{\hspace{-0.5em}*}\,r\,R_{n\ell}\,r^2\,dr,
\end{equation} 
which can be evaluated exactly for any $n$, $\ell$, and $\ell'$ to determine the photoionization cross-section as a function of energy.  We have calculated this cross-section from a variety of initial states and show them in the following section, demonstrating agreement with the result from box normalization calculations described below.

\section{Cross-sections for bound-to-continuum transitions: Box normalization}\label{sec:wellH}
A second approach commonly proposed to resolve the mathematical dilemma described in Section~\ref{sec:intro} is to place the system in a box of finite size, so that the problematic terms are neither zero nor infinite.\cite{orear}  We now demonstrate this approach, showing that the results converge to the correct values as the volume tends to infinity.

The potential energy of a hydrogen atom contained in a spherical well of radius $r_0$ with origin located at the atom's center of mass is 
\begin{equation}
 U\fbr{r}=
\begin{cases}
 -\frac{2\mathrm{Ry}\,\mathrm{a}_0}{r} & r < r_0 \\
 \infty & r \ge r_0.
\end{cases}
\end{equation}

We start by solving the Schr\"odinger equation to find the allowed energies $W_{n\ell}\fbr{r_0}$ and energy eigenfunctions for such a confined hydrogen atom.  As the system remains spherically symmetric, the wavefunctions retain the standard $Y_\ell^m$ angular portion of the free hydrogen atom.  

\begin{figure}
 \includegraphics[width=\columnwidth]{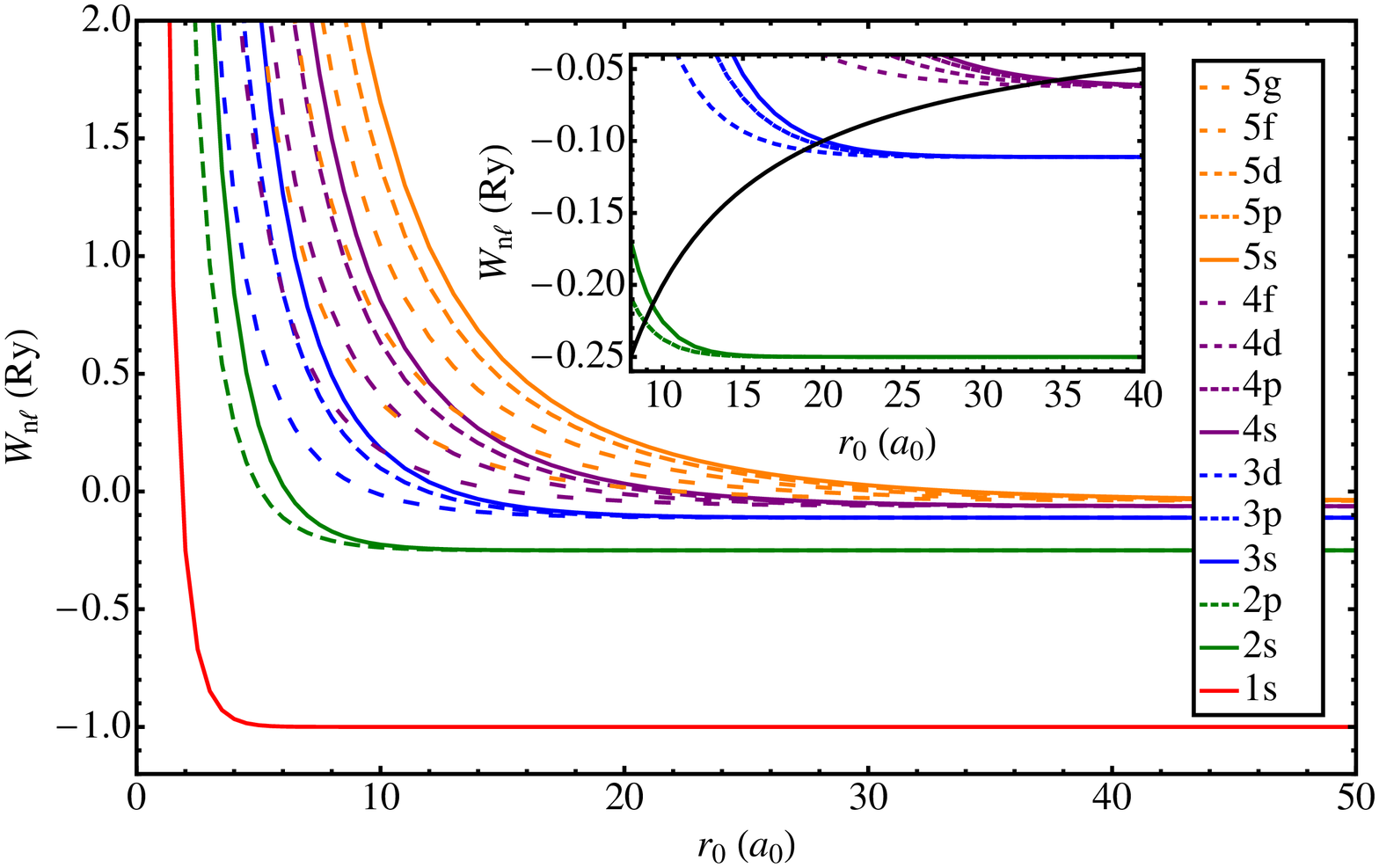}
 \caption{(color online) The energies of states as a function of radius of confinement.  The inset shows an additional curve representing the potential energy as a function  of $r$ on the same scale.}\label{fig:energies}
\end{figure}

We adopt the functional form of Eq.~\eqref{eq:normhydrogenabove} for each confined radial wavefunction, $\rconf_{W_{n\ell}\ell}$, and introduce the boundary condition:
\begin{equation}\label{eq:boundarycondition}
 \rconf_{W_{n\ell}\ell}\fbr{r_{\scriptscriptstyle 0}}\equiv 0.
\end{equation}
Allowed energies under this condition were found numerically for fixed $\ell$ and $r_0$ using Mathematica, and agree with those found previously by Aquino et al.\cite{aquino}
The boundary condition represented by Eq.~(\ref{eq:boundarycondition}) has the effect of raising the energies from those of the free atom bound states, as shown in Figure \ref{fig:energies}. 
The inset illustrates that the energies begin to rise more dramatically when the radius of confinement, $r_0$, is such that the state no longer has a region of exponential decay.  States previously in the continuum no longer exist since there is no continuum; we call those bound states which rise to an energy greater than the ionization limit \emph{pseudocontinuum} states, and relabel their energies $W$, dropping the subscript $n\ell$ to distinguish them from those of the (Coulombically) bound states.

These bound and pseudocontinuum wavefunctions were then numerically space-normalized in the usual way.  We wish to find prefactors for each of the initial and final states which will have the property of causing the the radial matrix element to tend, as $r_0{\rightarrow}\infty$, to that given in Section~\ref{sec:btc} for the free atom.

We define the density of final states with a particular symmetry $\rho\br{W}$ for the confined atom to be $\sfrac{2}{\br{W^+{-}W^-}}$ where $W^+$ ($W^-$) is the energy of the adjacent state with the same symmetry above (below) $\left|\rconf_{W\ell}\right>$.  Thus, as $r_0$ goes to infinity so too will the density of states above the ionization limit.  Furthermore, from Section \ref{sec:btc} as $r_0\rightarrow\infty$, pseudocontinuum states $\rconf_{W\ell}\br{r}$ assume zero amplitude in such a way that
\begin{equation}
 \sqrt{\rho\br{W}}\rconf_{W\ell}\fbr{r}\rightarrow R_{W\ell}^\mathrm{\,en}\fbr{r},\\
\end{equation}
and for (Coulombically) bound states, trivially
\begin{equation}
 \rconf_{W_{n\ell}\ell}\fbr{r}\rightarrow R_{n\ell}\fbr{r},
\end{equation}
where in each case $\rconf$, the confined-atom wavefunction, is space-normalized, and $R_{W\ell}^\mathrm{\,en}$ and $R_{n\ell}$ are energy-normalized and space-normalized, respectively.  The free atom bound-to-continuum radial matrix element in Eq.~\eqref{eq:DnlWl} can therefore be readily found from
\begin{align}\label{eq:curlyD}
 \mathfrak{D}_{W_{n\ell}\ell\rightarrow W\ell'} &\equiv \int\limits_0^{r_0}\sqrt{\rho\br{W}}\rconf_{W\ell'}^*\, r\mathfrak{R}_{W_{n\ell}\ell}\,r^2\,dr\\
 &\xrightarrow{r_0\rightarrow\infty} D_{n\ell\rightarrow W\ell'}.
\end{align}

\begin{figure}
 \includegraphics[width=\columnwidth]{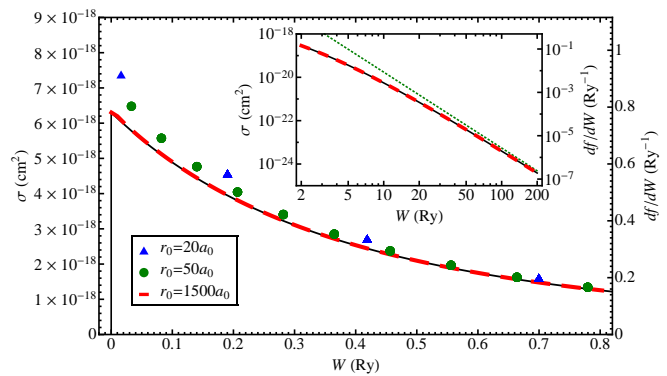}
 \caption{(color online) The photoionization cross-section for \mbox{$1\mathrm{s}\rightarrow W\!\mathrm{p}$.}  The results from the infinite well approach are shown with triangles for \mbox{$r_0=20\mathrm{a}_0$,} circles for \mbox{$r_0=50\mathrm{a}_0$,} and a dashed line for \mbox{$r_0=1500\mathrm{a}_0$,} and the solid line is the exact calculation.  The last two are shown in the inset for large energies, along with the dotted line showing the Born approximation.}
 \label{fig:oscstrengthplot1sWp}
\end{figure}

We have calculated the photoionization cross-sections for the hydrogen atom using the infinite well approach for \mbox{$1\mathrm{s}\rightarrow W\!\mathrm{p}$} for a variety of values of $r_0$ and show in Fig.~\ref{fig:oscstrengthplot1sWp}
that it converges towards those found using exact calculations from Section~\ref{sec:btc}, which agree with those from Ref.~\citenum{fano}.
The data at $r_0=20\mathrm{a}_0$ and $r_0=50\mathrm{a}_0$ is sparse as the densities of states at these radii of confinement are quite low; all allowed final energies in the range shown are plotted.  With $r_0=1500\mathrm{a}_0$, there are sufficiently many states in the energy range shown to produce a smooth curve on the scale of the figure.  The consistency with Fermi's Golden Rule is now apparent, in which the density of states evolves from an inverse energy difference: multiplication of the square of the matrix element by the density of states is mathematically equivalent to instead using an en\mbox{ergy- ra}ther than \mbox{space-nor}malized continuum wavefunction.  It can be seen that the cross-section converges to the true continuum case from above as $r_0$ increases.  This can be attributed to the fact that while $r\rconf_{W\ell}$ tends to a constant amplitude as $r\rightarrow\infty$, it does so from below; thus at low $r$ the amplitude is slightly reduced from this value.  Therefore when the wavefunction is truncated at finite $r_0$ before reaching its final amplitude, $\rconf_{W\ell}^2$ decreases disproportionately to increases in $r_0$.  This results in the $\mathfrak{D}_{W_{n\ell}\rightarrow W\ell'}$ of Eq.~\eqref{eq:curlyD} taking slightly too large a value for finite $r_0$.  The inset of Fig.~\ref{fig:oscstrengthplot1sWp} shows that this calculation works for a large range of final-state energies, over the entire non-relativistic regime.  At high energies, the cross section can be seen to approach that predicted by the Born approximation,\cite{fano} which for a transition $n\mathrm{s}\rightarrow W\mathrm{p}$ takes the form:
\begin{equation}
 \sigma\xrightarrow{W\rightarrow\infty}\frac{2^9\pi \mathrm{a}_0^3 \mathrm{Ry}^{9/2}}{3c\hbar n^3}W^{-\sfrac{7}{2}}.
\end{equation}

An alternative measure of the likelihood of a bound-to-continuum transition is the oscillator strength distribution,\cite{gallagher} $\frac{df}{dW}$, which is related to the cross-section, by
\begin{equation}
 \frac{df}{dW}=\frac{c\hbar}{8\mathrm{a}_0^3\mathrm{Ry}^2\pi^2}\sigma ,
\end{equation}
and for completeness these values are given in the figures as well.

\begin{figure}
 \includegraphics[width=\columnwidth]{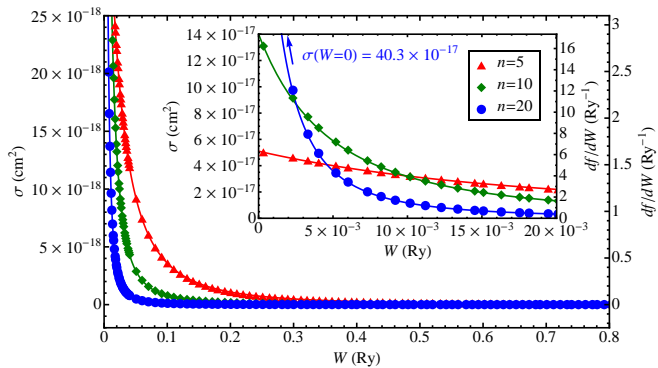}
 \caption{(color online) The photoionization cross-section for \mbox{$\left\lbrace 5,\,10,\,20\right\rbrace\mathrm{s}\rightarrow W\!\mathrm{p}$.}  The exact calculations are shown as lines, with symbols showing results using the infinite well approach with \mbox{$r_0=1500\mathrm{a}_0$}. $\mathrm{Ry}$ is the Rydberg unit of energy.}
 \label{fig:crossSection-nsWp}
\end{figure}
We also show in Fig.~\ref{fig:crossSection-nsWp} a plot to illustrate that this approach works for a range of values of initial $n$.  Higher $n$ results in a greater cross-section for low ejection energies, and a lesser cross-section for high ejection energies.  This relates to the general observation\cite{merkt} that high $n$ (Rydberg) states are strong absorbers of long-wavelength radiation.  The peak cross section occurs at threshold; for $n\mathrm{s}\rightarrow\br{W{=}0}\!\mathrm{p}$, this simplifies to
\begin{equation}
 \sigma=\frac{2^7n^5\pi^2\mathrm{a}_0^2}{411}\!\br{\sum\limits_{m=0}^{\infty}\!\frac{2^m\br{1{-}n}^{\br{m}} L_{-5-m}^{\br{3}}\fbr{-2n}}{m!}}^2,
\end{equation}
where $L_n^{\br{\alpha}}\br{x}$ is a generalized Laguerre polynomial, and $\br{1-n}^{\br{m}}=\br{1-n}\br{2-n}\cdots\br{m-n}$ is the Pochhammer function, which becomes zero for any $m\ge n$, forcing the sum to terminate.

\section{Conclusion}
We have presented two methods for calculating bound-to-continuum cross-sections: the first employing energy-normalization and the second, box-normalization, and demonstrated convergence between them, using photoionization of the hydrogen atom as a concrete example.  We thereby elucidate some quantitative aspects of Fermi's Golden Rule and show the product of the density of states and a space-normalized continuum wavefunction to be mathematically equivalent to using only an energy-normalized continuum wavefunction.

\begin{acknowledgments}
The authors wish to acknowledge NSERC for financial support.  We thank Bill Atkinson for critical reading of the manuscript, and Eric Brown and Jaclyn Semple for preliminary discussions.
\end{acknowledgments}

\bibliography{Shiell_Ugray_AJP}{}

\begin{thebibliography}{}
\bibitem{wavepacket}
Note that for more general electromagnetic wave packets, a full-time-dependent
  analysis is required; see Refs.~\citenum{seideman,yeazell} for more details.
\end{thebibliography}
\bibliographystyle{ajp}

\end{document}